\documentclass[aps,amsmath,amssymb,reprint,showkeys]{revtex4-2}

\usepackage[dvipsnames]{xcolor}
\usepackage{graphicx}
\usepackage{hyperref}
\usepackage[utf8]{inputenc}
\usepackage[T1]{fontenc}
\usepackage{physics}
\usepackage{mathtools}
\usepackage{siunitx}
\usepackage{bm}
\usepackage{dcolumn}
\usepackage{appendix}
\usepackage{lipsum}
\usepackage{multirow}
\usepackage{chemformula}

\usepackage[normalem]{ulem}

\begin{document}

\title{1D Kinetic Energy Density Functional Learned with Symbolic Regression}
\author{Michael A.J. Mitchell}\email{mitchemi@tcd.ie}
\author{Teresa Del Aguila Ferrandis}
\author{Stefano~Sanvito}
\affiliation{School of Physics, AMBER and CRANN Institute, Trinity College, Dublin 2, Ireland}
\date{\today}

\begin{abstract}
Orbital-free density functional theory promises to deliver linear-scaling electronic 
structure calculations. This requires the knowledge of the non-interacting kinetic-energy 
density functional (KEDF), which should be accurate and must admit accurate functional 
derivatives, so that a minimization procedure can be designed. In this work, symbolic regression 
is explored as an alternative means to machine-learn the KEDF, which results into analytical 
expressions, whose functional derivatives are easy to compute. The so-determined semi-local 
functional forms are investigated as a function of the electron number, and we are able to track 
the transition from the von von Weizs\"acker functional, exact for the one-electron case, to the 
Thomas-Fermi functional, exact in the homogeneous electron gas limit. A number of separate 
searches are performed, ranging from totally unconstrained to constrained in the form of an 
enhancement factor. This work highlights the complexity in constructing semi-local approximations 
of the KEDF and the potential of symbolic regression to advance the search.
\end{abstract}

\maketitle

\section{Introduction}
The most widely used approach to \textit{ab initio} electronic structure calculations 
is density functional theory (DFT)~\cite{Parr1995}, which is typically chosen for its 
favourable balance between theoretical rigor, computational throughput and predictive
accuracy~\cite{Jones2015_DFT_Success_Story,Marzari2021_Electronic_Structure_Methods_For_Materials_Design}. 
The Hohenberg-Kohn (HK) theorems establish a one-to-one relation
between the ground-state energy, $E_0$, and the ground-state electron 
density, $\rho_0$, of a given system of atoms, molecules, or condensed 
phases~[\citenum{Hohenberg1964_HK_Theorem}]. The Rayleigh-Ritz variational formulation 
of DFT~[\citenum{Kvaal2015GroundstateDF}], 
\begin{equation}\label{Eq1}
    \min_\rho \{\mathcal{E}[\rho] + E_\text{ext}[\rho]\}=\mathcal{E}[\rho_0] + 
    E_\text{ext}[\rho_0]=E_0\:,
\end{equation}
effectively offers a computational implementation of the HK theorem. Here, the
ground-state energy is found by minimizing the universal DFT energy functional, 
$\mathcal{E}[\rho]$, together with the energy associated to an external potential
$v_{\text{ext}}$, namely $E_\text{ext}[\rho]$, where the minimization is performed over 
all the relevant $N$-particle electron densities, $\rho$. The universal DFT functional,
then comprises of two terms, the total kinetic-energy functional, $T[\rho]$, 
and the electron-electron Coulomb interaction energy functional, $U_\text{ee}[\rho]$. 
As it stands, unfortunately, the problem cannot be solved, since both $T[\rho]$
and $U_\text{ee}[\rho]$ are unknown.

The standard procedure to tackle the problem is then to single out from the kinetic 
energy its non-interacting component, $T_\text{s}[\rho]$, and from the Coulomb 
one the classical Hartree part, $E_\text{H}[\rho]$, leaving the remaining 
contributions bundled in the so-called exchange-correlation energy, 
$E_\text{xc}[\rho]$. Although, $E_\text{xc}[\rho]$ can be further separated 
into an exchange and a correlation part, $E_\text{x}[\rho]$ and $E_\text{c}[\rho]$
respectively, it is often approximated as a whole, since one may benefit from 
favorable error cancellations~\cite{Dreizler1990}. Following this procedure the 
universal DFT functional writes 
\begin{equation}\label{Eq2}
\mathcal{E}[\rho]= T_\text{s}[\rho] + E_\text{H}[\rho] + E_\text{xc}[\rho]\:.
\end{equation}
Interestingly, our knowledge of these three terms varies. The Hartree energy is 
just the classical Coulomb repulsion associated to the electron charge density and 
it is a well-defined functional of $\rho$. In contrast, the exact form of 
$E_\text{xc}[\rho]$ is unknown, but there exists a vast range of approximations 
constructed from the density, its gradient or higher derivatives~\cite{LibXC}. 
Finally, the non-interacting kinetic energy is easily calculated from the single
particle wave-function, but a functional form in terms of the density is unknown, except
for a few limiting cases. 

Because of this last restriction, the minimization of the total energy functional
cannot simply proceed by numerically implementing Eq.~(\ref{Eq1}), and the Kohn-Sham (KS)
construct~\cite{Kohn1965_DFT_Origin_Paper} is then the only choice. In contrast, if
one insists in performing a direct minimization of the energy functional, an approach
known as orbital-free DFT (OFDFT)~\cite{Mi2023OrbitalFreeDF}, then approximations to
the non-interacting kinetic-energy density functional (KEDF) must be considered.
Unfortunately, despite many years of research, accurate universal approximations 
are not available and OFDFT remains confined to be used only for specific materials
systems~\cite{Mi2023OrbitalFreeDF,Xu_Qiang_Ma_2024,KARASIEV_2012_Issues_Challenges_OFDFT}. 

In recent years, machine learning (ML) has been proposed as a tool to numerically
construct approximations to the KEDF
\cite{Snyder_2012,Seino2018,Mazo-Sevillano2023,Sun2024,Zhang2024}.
Since the training set needed for learning the KEDF consists of solutions of 
non-interacting problems with different external potentials, this appears as a 
rather intriguing and computational undemanding task. However, as noticed early
in~\cite{Snyder2015,Li2016}, although accurate numerical ML KEDFs can be 
constructed in a relatively easy way, the access to accurate functional 
derivatives is significantly more demanding. This, then, imposes different 
minimization strategies and the need to learn, together with the KEDF, its 
functional derivatives~\cite{Meyer2020,Imoto2021,Ryczko2022}.

In an attempt to overcome these difficulties here we explore a different
strategy, namely we attempt at defining analytical KEDFs by using symbolic
regression. In this case, both the functional and its derivatives are explicitly
known and no issues concerning the density minimization arise. 
In particular, we focus on the simple one-dimensional case and 
consider local functionals constructed from the electron density and
its derivatives. For one electron, we find that the symbolic regression is 
always able to converge to the exact solution, namely the von Weizs\"acker 
functional. In contrast, a semi-local expression of the functional appears not 
sufficient, in the few-electron case. 

The paper is organized as follows. In the next section we first introduce the 
foundational concepts of OFDFT and then we discuss our symbolic regression
algorithm and the modifications made to deal with functionals and not just
functions. The section is completed with a discussion of the specific model
used here and the construction of the dataset. Then, we move to the results 
part, presenting three different search strategies, namely fully unconstrained,
informed with the von Weizs\"acker and Thomas-Fermi functionals, and 
a search for the Thomas-Fermi enhancement factor. Finally, we conclude.

\section{Problem statement and method}
\subsection{Orbital Free Density Functional Theory}
In this section we overview the basic concepts of OFDFT and we discuss the
various classes of KEDFs that will be considered. Our starting point is the
universal DFT functional, defined in Eq.~(\ref{Eq2}). Since a density functional 
form for $T_\text{s}[\rho]$ is not available, the energy functional is usually 
minimized using the KS scheme. The idea is to map the fully interacting problem 
onto a non-interacting one sharing the same ground-state electron density and 
energy. This is achieved through the introduction of a ``fictitious'' one-particle
potential, the KS potential $v_\text{KS}(\mathbf{r})$, which defines a set of
Schr\"odinger-like equations
\begin{equation}\label{Eq3}
\left[-\frac{1}{2}\nabla^2 + v_{\text{KS}}(\mathbf{r})\right]\phi_i (\mathbf{r})=
\epsilon_i \phi_i (\mathbf{r})\:.
\end{equation}
In Eq.~(\ref{Eq3}) the $\phi_i(\mathbf{r})$ are the fictitious non-interacting 
KS orbitals and the $\epsilon_i$ the associated orbital energies. Strictly
speaking these are not interpretable as physical single-particle states and the 
energies cannot be associated to removal energies~\cite{Perdew1983}.

Nevertheless, the KS orbitals allow us to compute the kinetic energy in the
absence of a density functional, namely we can write,
\begin{equation}\label{Eq:KEDLap}
    T_\text{s}[\rho] = - \frac{1}{2}\int_{\mathbb{R}^3}\sum_{i=1}^\mathrm{occ} \phi_i^*(\mathbf{r})\nabla^2 \phi_i(\mathbf{r}) d\mathbf{r}
    =\int_{\mathbb{R}^3}\tau_\text{s}^{\text{I}}(\mathbf{r})d\mathbf{r},
\end{equation}
where the sum runs over the occupied states (``occ'') and natural units,
$\hbar=m=1$, are considered. Equation (\ref{Eq:KEDLap}) also introduces
the kinetic-energy density $\tau_\text{s}^{\text{I}}(\mathbf{r})$. 
Knowing the kinetic energy exactly is a key aspect of the KS scheme, since
$T_\text{s}[\rho]$ is of the same order of magnitude of the total energy 
\cite{Xu_Qiang_Ma_2024,KARASIEV_2012_Issues_Challenges_OFDFT}.
Note also that we can locally transform the kinetic-energy density,
$\tau_\text{s}^{\text{II}}(\mathbf{r}) := \tau_\text{s}^\text{I}(\mathbf{r}) 
- \frac{1}{4}\nabla^2 \rho(\mathbf{r})$, without changing the kinetic energy, 
namely
\begin{equation}\label{Eq:KEDPositive}
    T_\text{s}[\rho] = \int_{\mathbb{R}^3} \tau_\text{s}^\text{II}(\mathbf{r})d\mathbf{r}=
     \frac{1}{2}\int_{\mathbb{R}^3} \sum_{i=1}^N |\nabla \phi_i(\mathbf{r})|^2\:, 
\end{equation}
where $\rho(\mathbf{r})=\sum_{i=1}^\mathrm{occ}|\phi_i(\mathbf{r})|$.

As mentioned before, the KEDFs in Eq.~(\ref{Eq:KEDLap}) and Eq.~(\ref{Eq:KEDPositive}) 
are not explicit functionals of the electron density, a situation encountered also for 
orbital-dependent approximations of the exchange-correlation 
functional~\cite{Gill_Johnson_Pople_Frisch_1992}. In this case, typical KS DFT 
calculations scale cubically with the size of the basis set, since a digonalization
step is required at every self-consistent iteration of the KS 
equations~\cite{Demmel1997_Applied_Numerical_Linear_Algebra}. Alternatively, 
one might attempt to find a unitary transformation that orthogonalises the 
eigenstates, although this also has cubic scaling~\cite{Parr1995}. Intriguingly,
there is no fundamental reason to suggest that DFT predictions \textit{must} display 
cubic scaling. The principle of near-sightedness~\cite{Kohn1978_Locality_Principle},
in fact, indicates quite the opposite, namely that cubic scaling is avoidable. 
This establishes that the spatial correlation in systems with localised electrons 
drops off exponentially, meaning that it is unnecessary to invest computational 
resources in calculating contributions to the energy far beyond the local 
environment. This intuition is partially evidenced by the recent successes of 
linear-scaling machine-learning-based force fields, wherein locality is 
encoded as an inductive bias via the use of a cut-off radius~\cite{unke_machine_2021}.

The variational approach to DFT suggests how the KS construction may be, in principle,
avoided. In fact, one may minimize the Lagrangian,
\begin{equation}
    \mathcal{L}[\rho] = \mathcal{E}[\rho] + \int d\mathbf{r} \rho(\mathbf{r}) v_\text{ext}(\mathbf{r}) - \mu\left[\int d \mathbf{r}\rho(\mathbf{r}) - N\right],
\end{equation}
where the chemical potential Lagrange multiplier, $\mu$, ensures the correct number 
of electrons, $N$. The minimization may be performed by demanding that the 
functional derivative of $\mathcal{L}[\rho]$ with respect to the density vanishes,
a condition that leads to the equality
\begin{align}
    \mu &= \frac{\delta T_\text{s}[\rho]}{\delta\rho(\mathbf{r})} + \left[\frac{\delta E_\text{H}[\rho]}{\delta\rho(\mathbf{r})} + \frac{\delta E_{\text{xc}}[\rho]}{\delta \rho(\mathbf{r})} + v_\text{ext}(\mathbf{r})\right], \label{Eq:FullLagrange}\\
    &\equiv \frac{\delta T_\text{s} [\rho]}{\delta \rho(\mathbf{r})} + v_\text{KS}(\mathbf{r}).\label{Eq:KSLagrange}
\end{align}
Assuming $E_\text{xc}[\rho]$ is well approximated, the search of the ground-state 
energy, $E_0[\rho_0]$, at the minimum of $\mathcal{L}[\rho]$ depends entirely
on the knowledge of $T_\text{s}[\rho]$. Should $T_\text{s}[\rho]$ admit local or
semi-local approximations, real-space, linearly-scaling, parallelisable
minimisation procedures will be possible \cite{Prentice2020TheOL}. This consideration 
is at the heart of OFDFT.

The most elementary local and semi-local approximations to the KEDF are the Thomas 
Fermi (TF) functional \cite{Thomas_1927_TF,Fermi_1927_TF,Dirac_1930_TFD}, 
\begin{equation}
T_\text{TF}[\rho] = \frac{3h^2}{40m_e}\left(\frac{3}{\pi}\right)^{\frac{2}{3}}\int \rho(\mathbf{r})^{5/3}d\mathbf{r},
\end{equation}
and the von Weizs\"acker (vW) functional \cite{Weizscker1935ZurTD},
\begin{equation}
    T_\text{vW}[\rho]=\frac{\hbar^2}{8m_e} \int \frac{|\nabla \rho(\mathbf{r})|^2}{\rho(\mathbf{r})}d\mathbf{r}\:,
\end{equation}
where $m_e$ is the electron mass and $\hbar=h/2\pi$ the reduced Plack constant.
These two functionals occupy a particular place, since they are exact, respectively
in the limit of the non-interacting homogeneous electron gas (TF) and in that of a
single electron problem (vW). Unfortunately, their accuracy in a general case (for a 
generic external potential and electron number) is rather low and both or any linear 
combination of them fail to produce the shell structure of atoms~\cite{Yonei1965OnTW}. 

Beyond these two functionals the gradient-expansion approximation
\cite{Kirzhnits1957_Quantum_Correction_TF} improves on $T_\text{TF}[\rho]$ by 
incorporating some density gradient information, but it diverges at the Kohn
singularity \cite{Kohn1959_Image_Fermi_Surface_Vibration_Specturm_Metal}. 
Functionals belonging to the generalised gradient approximation (GGA) 
\cite{Perdew1996GeneralizedGA} and meta GGA (mGGA) type, containing higher-order 
derivatives of the density, have also been developed. These are strongly informed 
by the slowly varying homogeneous electron gas \cite{Murphy1979GradientEO} and 
exchange-correlation functionals \cite{Lee1991ConjointGC}, and do recover the 
lower order gradient-expansion approximation results, while mitigating the 
divergences. Further improvements have to abandon the local nature of the
functional and typically have a two-point form. This class encodes more directly the 
non-linear structure of the KEDF 
\cite{Wang1992KineticenergyFO,Smargiassi1994OrbitalfreeKF,PhysRevB.58.13465}, 
but it will not be investigated here. All functionals under consideration must 
obey the proper KEDF scaling laws \cite{PhysRevA.32.2010}, namely where 
$\rho_\lambda (\mathbf{r}) := \lambda^3 \rho(\lambda\mathbf{r})$, for $\lambda$ 
being a positive real constant, we must have
\begin{equation}
    T_\text{s}[\rho_\lambda] = \lambda^3 T_\text{s}[\rho]\:. \label{eq:Scaling}
\end{equation}

Despite the strong incentive for finding a KEDF to facilitate OFDFT calculations, 
their predictive accuracy is simply not satisfactory yet for complex chemical systems 
\cite{Xu_Qiang_Ma_2024}. With the KE contribution to the KS Hamiltonian being the 
largest, approximate KEDFs have an extremely tight margin of error for remaining 
practically useful. The difficulty of this task is compounded by the 
unavailability of non-local pseudopotentials in the orbital-free 
context \cite{Cabral_Martins_1995,Watson_Jesson_Carter_Madden_1998}, though 
we will not address such problem here. 

Finally, besides the difficulty of finding an accurate kinetic-energy functional, another 
pressing issue in the development of KEDFs is achieving accuracy in the functional derivative, 
an obstacle that is especially relevant when constructing the functional from machine learning 
\cite{Snyder_2012}. The issue is that usually the functional derivative of a many-parameter 
model can become wildly inaccurate. This is due to the curse of dimensionality 
\cite{Bronstein2021GeometricDL,Pokharel2022ExactCA}, which prevents
the extrapolation (and hence the transferability) capabilities of any many-parameter model. 
In this case machine-learning KEDFs have little knowledge beyond the ground-state 
densities included in the dataset, such as those encountered during the course of a self-consistent 
cycle. In this work we will not be focusing on the functional derivative 
itself, but on symbolic regression as a technique to deduce simpler, analytic functionals. 
These are better posed to avoid issues brought on by model over-complexity
\cite{Hernandez2022GeneralizabilityOF,Ma2022}. With explicit expressions for
the functional, it becomes easier to confirm and build them in adherence to analytic 
constraints~\cite{Pokharel2022ExactCA}.

\subsection{Symbolic Regression}
In this work we explore Symbolic Regression (SR) as a technique for generating 
interpretable KEDFs. SR is a data-driven approach to discovering explicit expressions to fit 
data following the Keplerian trial-and-error paradigm \cite{Makke2022InterpretableSD}. 
SR algorithms traditionally employ genetic programming as a means 
to perform the discrete optimisation over the space of analytic expressions up to a given 
complexity. Despite the relatively sluggish convergence rate of genetic programming in 
the absence of gradient descent, faced with the combinatorially enormous discrete space 
of possible expressions \cite{Yuan2017IdentifyingMO}, there is not much choice besides 
genetic programming or random search. This is reflected in the recent result showing that 
SR is a NP-hard problem \cite{Antonov2024AFA,Virgolin2022SymbolicRI}. 
However, the risk of limited model flexibility and expressivity when compared 
to deep learning approaches is mitigated by the simplicity of SR, which is less prone to 
overfitting and is poised to generalise better \cite{Hernandez2019FastAA,Angelis2023ArtificialII}. 
SR is a multi-objective optimisation over the predictive error and complexity; an ``Occam's razor'' 
of sort \cite{Wang2019SymbolicRI}. As it is a genetic-algorithm technique, typically the Pareto 
front analysis involves identifying the largest reduction in the prediction error for the smallest increase 
in model complexity (see Section~\ref{Results}). 
\begin{figure}[h]
    \centering
    \includegraphics[width=0.41\textwidth]{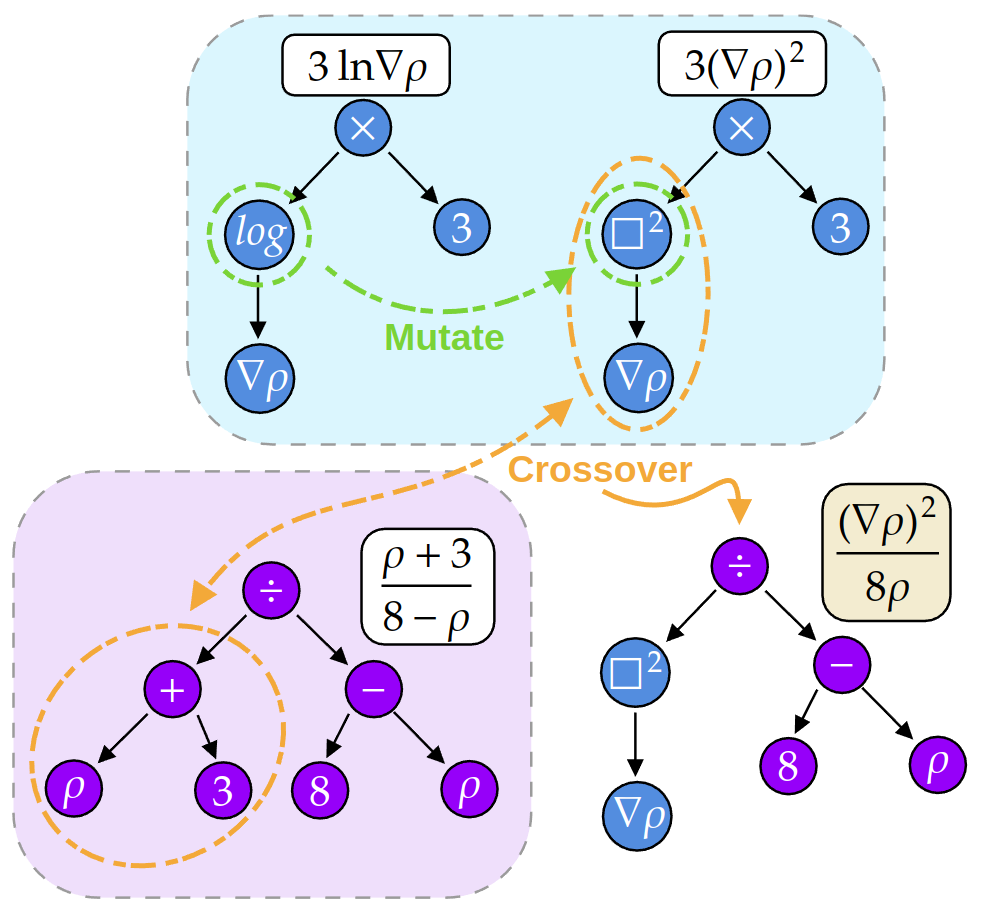}
    \caption{Schematic representation of the symbolic regression process. A diagram depicting a mutation, 
    $3\ln \nabla \rho \to 3(\nabla \rho)^2$, and a crossover operation,
    $\left[3(\nabla \rho)^2,\ \frac{\rho + 3}{8 - \rho}\right] \to \frac{(\nabla \rho)^2}{8\rho}$, in the tree-structure representation 
    of symbolic expressions.}
    \label{fig:trees}
\end{figure}

Despite recent advances of large-language-model-based SR algorithms 
\cite{Udrescu2019AIFA,Holt2023DeepGS,Kamienny2022EndtoendSR}, the state of the art at the time of 
commencing the project was the PySR library \cite{Cranmer2023InterpretableML}, and more specifically 
the SymbolicRegression.jl component. While, to the best of our knowledge, there are no recent papers  
assessing whether PySR is still state of the art, it is clear that genetic-programming-based SR algorithms are 
strongly competitive~\cite{Radwan2024ACO}.
PySR employs genetic-programming search together with simulated annealing, a combination that essentially
smoothly transforms the algorithm from a random search (high temperature) to a hill-climbing optimisation 
(low temperature) \cite{Kantor2021SimulatedAF}. We have modified PySR to deal directly with functionals
and not just with functions. A high-level overview of the algorithm is as follows:
\begin{itemize}
    \item Equations are represented as trees, composed of the allowed operators, input data, and equation constants [see examples 
    in Fig. (\ref{fig:trees})]. Each equation tree constitutes an ``individual'' in the context of a genetic algorithm.
    \item Individuals are organised into separate populations, which evolve independently to take advantage of parallelisation.
    \item Within a given population, a ``tournament'' is performed. This involves selecting a subgroup of the population, subjecting the 
    equations' evaluation over the input data to the loss function, and selecting the fittest individual with probability $p$, the second 
    fittest individual with probability $p(1-p)$, and so on.
    \item Given this selected individual, make copies and apply mutations to the underlying tree structure of the equation.
    \item Replace the weakest (or oldest) members of the population with the mutated individuals.
    \item Perform asynchronous migration between the populations with a certain probability.
    \item Introduce a notion of temperature, $T \in [0, 1]$ such that simulated annealing may be employed via 
    $p = \exp \left(\frac{L_F - L_E}{\alpha T}\right)$ on the tournament selection step, where $L_F$, $L_E$ are 
    the fitness of the mutated and original individuals respectively, and $\alpha$ is a hyperparameter. As $T$ decreases 
    during the course of the search, the selection criteria becomes less and less permissive of accepting equations, which 
    worsen performance on the loss function. 
\end{itemize}

\subsection{SymbolicRegression.jl Adaptation}
\begin{figure}[h]
    \centering
    \includegraphics[width=0.4\textwidth]{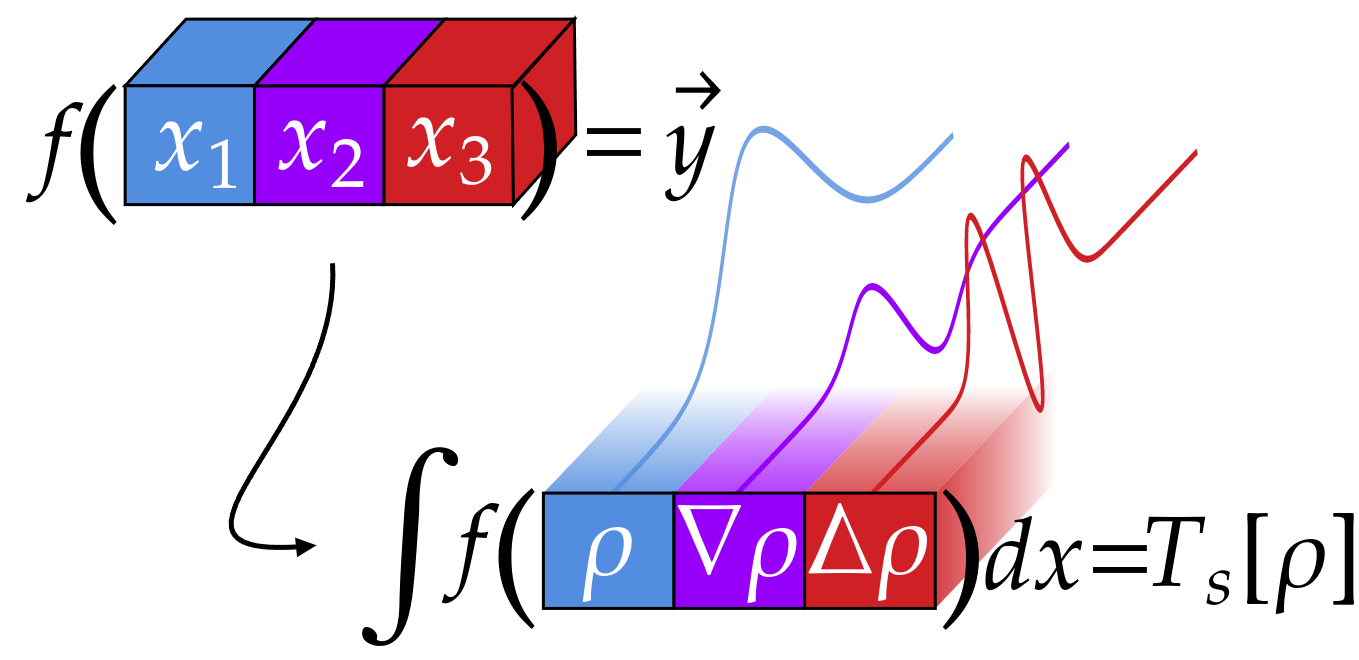}
    \caption{A diagram depicting the modification required by the SymbolicRegression.jl library to 
    work with functionals. The density, and higher order derivatives of the density are passed as 
    features for the programme to search over. The resulting expression is then treated as a 
    integrand, and the result of the integral is passed to the loss function.}
    \label{fig:SRjlmodification}
\end{figure}
SR has been featuring more routinely in other areas of materials science 
\cite{He2021MachineLA,Wang2019SymbolicRI,Hernandez2022GeneralizabilityOF,Burlacu2023}, and some use of SR in 
the optimization of exchange-correlation functionals has also been documented~\cite{Ma2022}. However, to the best of our 
knowledge, the same has never been attempted for the KEDF. The SymbolicRegression.jl library has been developed for finding 
low-dimensional functions over high-dimensional datasets, rather than for finding functionals, and for this reason it has been 
modified here. SymbolicRegression.jl anticipates a datapoint in a dataset to be a $D$ dimensional input vector, with one 
dimension for each input feature. We use this axis of the input vector to refer to different functions, $\rho(x),\ \frac{d}{dx}\rho(x)$, and 
other constructions involving the electron density and its derivative, for example the TF and vW kinetic-energy densities. However, in 
order to carry 1D spatial information relating to these functions/densities, an additional dimension must be added to the input data. 
To facilitate this extra dimension, the library has been edited to accommodate the additional memory allocations needed, along with 
the promotion of all operators to point-wise matrix operators. The loss function also now involves an integral over the candidate
kinetic-energy densities, say $f(x)$, yielding a value, which is compared against the total, known $T_\text{s}[\rho(x)]$ for that $\rho(x)$
[see Fig. (\ref{fig:SRjlmodification})], as calculated by direct diagonalisation. As standard in an ML setting, the mean-square error (MSE)
is taken across the dataset. This modification lets us take advantage of the excellent parallelism capabilities of the library.

\subsection{Description of model and associated dataset}
\begin{figure*}
    \centering
    \includegraphics[width=\linewidth]{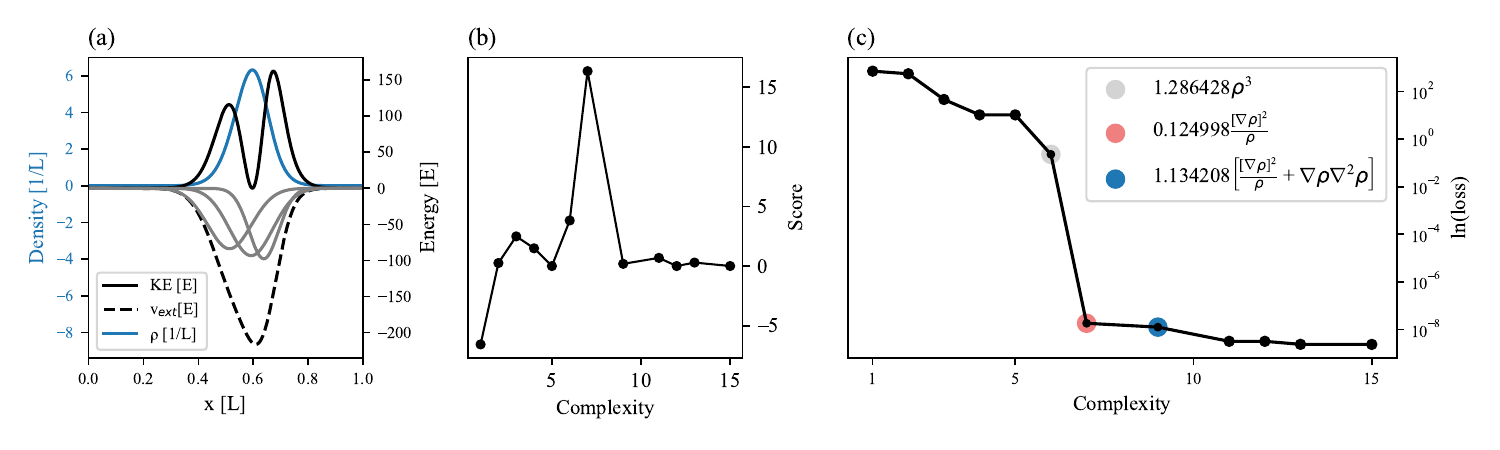}
    \caption{An example of the symbolic regression procedure for the case of one electron in its ground state.
    In panel (a) we show the ground-state density (solid blue line) and kinetic-energy density (solid black line) 
    for one electron in the external potential (dashed black line) made of three overlapping Gaussians (grey solid 
    lines). Panel (b) displays the score of different kinetic-energy densities against their complexity. The score 
    metric is a combined measure of predictive performance and expression complexity level, with each arithmetic 
    operation contributing 1 to the complexity. Panel (c) presents the Pareto front of the optimization procedure. 
    This is defined at the steepest drop in the loss function between consecutive equations of increasing complexity. 
    In this particular case, the Pareto front is found at the outset of the vW functional. 
    }
    \label{fig:Pareto}
\end{figure*}
In this work we consider a simple one-dimensional (1D) problem, which is numerically lighter than working in 
higher dimensions. In fact, one has to consider that genetic searches are computationally expensive, since
a large number of iterations are usually performed. Furthermore, in our case we are dealing with functionals,
meaning that millions of numerical integrations must be performed. A one-dimensional setup is also
numerically convenient to generate easily the exact ground-state energies and densities to be included in the
training set. In particular, we perform exact diagonalization calculations in the $[0,1]$ interval segmented 
over a uniform grid of $N=2,000$ points, $x_i$. Each system is enclosed in an infinite potential well, enforced via 
Born-von Karman boundary conditions \(\psi(x_0) = \psi(x_{N-1}) = 0\).

The data for the machine learning task is computed from 1D systems comprising a number of non-interacting, 
spinless electrons, $N_e$, ranging from 1 to 20. The external potential is constructed from three superimposed 
Gaussian distributions, 
\begin{align}\label{Gauss}
v_\text{ext}(x) = - \sum_{i = 1}^{3}A_i e^{(x-b_i)^2/(2c_i^2)}\:,
\end{align}
which are visualised as the grey dotted lines in Fig.~\ref{fig:Pareto}. The Hartree and exchange-correlation 
terms are not included, since we are searching for the KEDF, and we always consider ground-state solutions.
The parameters \(A_i\), \(b_i\), and \(c_i\), defining the Gaussian potential are chosen randomly for each system, 
within a defined range. In order to avoid interactions with the infinite well boundary, the \(A_i\) parameter range, 
dictating the depth of the wells, is defined according to $N_e$ in order to confine the electrons within the gaussian 
potential. The derivatives of the density are calculated numerically, using the central differences method except 
for the first and last grid points, where we use the forwards and backwards differences, respectively. The kinetic 
energy is calculated using the one-dimensional version of the positive definite KEDF, $\tau_\text{s}^\text{II}(x)$, of 
Eq.~(\ref{Eq:KEDPositive}). 

An example on how the symbolic regression progresses is provided in Fig.~\ref{fig:Pareto}. Here we consider 
the KEDF for a single electron in the external potential [see dashed black line in panel (a)] resulting from the 
superposition of three Gaussians (solid grey lines). The resulting charge density has a single-peak structure (solid 
blue line), while the associated KE density (solid black line) has a three-peak structure approximately mapping on 
the minima of the external potential. Panel (c) shows the progression of a symbolic regression as a function of 
complexity, which is defined as the number of operations needed to write a particular formula (more precisely
the number of nodes in an expression tree). As one can see from the picture, the Pareto front appears between 
complexity 6 and 7, when the algorithm moves from a kinetic-energy functional with cubic dependence on the 
density to the vW one. Note that additional complexity, leading to a correcting term to the vW functional, does 
not change significantly the loss. The exact position of the Pareto front can then be determined by the score 
of different evaluated kinetic-energy densities against their complexity [panel (b)]. The score, $S$, combines 
the expression loss, $L$, and the expression complexity, $C$, weighted by a parsimony term, $\lambda$, as 
$S = L + \lambda C$.
We will elaborate on the parsimony term as it pertains to our specific search later in the text. In the case of 
the example of Fig. \ref{fig:Pareto}, the overall best performing formula is obtained with a close approximation 
of the vW kinetic-energy density, which has complexity 7. 


\section{Results}\label{Results}
\subsection{Unconstrained Search}
Let us start our SR experiments by performing a completely unconstrained search, namely by searching for a
functional form of the KEDF, which is completely uninformed of any previous knowledge. In this case, we use 
the electron density, together with its first and second derivatives as input variables, and we allow the following
operations: addition, multiplication, division and squaring. No restriction is placed on the on the dimensionality 
of the functional and we perform SR for $N_e$ ranging from 1 to 20.
\begin{figure}[h]
    \includegraphics[scale = 0.8]{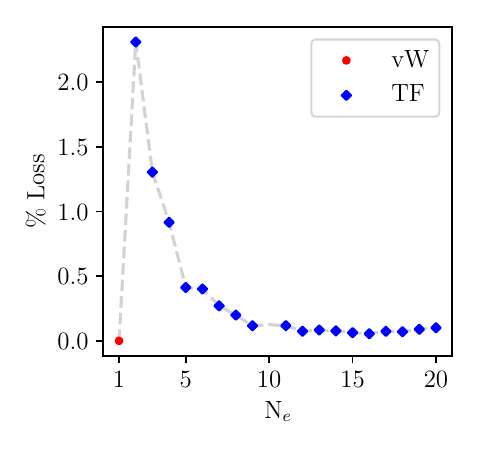}
    \caption{Percentage error of the best-performing functionals found by the SR against the total number of electrons, $N_e$. 
    In this case the search is unconstrained and unbiased and the input is the density and up to its second derivative.
    The vW functional is found for the single-electron case, while the TF one is the best approximation at any other 
    electron number.
    }
    \label{fig:UCBE}
\end{figure}

For the one-electron case ($N_e=1$), as expected, the SR search finds the vW functional to be the optimal one,
with a percentage loss of approximately 0.00027\% (see figure \ref{fig:UCBE}). The error is essentially associated 
to the multiplicative factor, which is estimated to be 0.124998460964301, instead of the exact value of 1/8. Moving 
away from the one-electron case, in any other situations the SR converges to some variants of the TF functional, 
which is associated to a kinetic-energy density of $\pi^2/6 \rho(x)^3$ in 1D \cite{Li2016}. As somehow expected, we 
find that the percentage loss decreases with increasing the number of electrons, namely as the system approaches 
the 1D homogeneous electron gas limit. In all cases, the computed numerical prefactor remains always below the
expected value of $\pi^{2}/6 \approx 1.6449$ and it seems to slowly converge to a value around $1.635$ for large 
$N_e$. This essentially means that the theoretical 1D TF functional overestimates the kinetic energy of multi-electron
systems with a non-uniform density resulting from the confining potential and the limited number of electrons.

Already in the physics-uninformed search, presented in Fig. \ref{fig:UCBE}, we have observed emerging a rather 
general trend. The single-electron and the many-electron systems are accurately captured by the familiar vW and TF 
functionals. The most difficult systems remain those with a few electrons, and in fact the largest percentage loss
is found for $N_e=2$. This is not surprising. By looking at the known expressions for the kinetic-energy density in term 
of single-particle orbitals [Eq.~(\ref{Eq:KEDLap}) or Eq.~(\ref{Eq:KEDPositive})], it is clear that the nodal structure of the
orbitals is crucial to determine the structure of the kinetic energy. Orbital orthogonality means that lower-lying orbitals 
dictate the nodal structure of the higher energy ones, being forced to interlace by Sturm's separation theorem \cite{teschlordinary}. 
This non-local structure underlying the density is intimately connected to the kinetic energy, meaning that the semi-local 
paradigm intrinsic to our approach is fundamentally limited. The inclusion of higher derivatives with the hope of 
``seeing further'' introduces greater sensitivity to numerical error. This renders the functional less practical for 
converging a calculation, so that we remain at the level of the second derivative.

\subsection{Search including vW and TF functionals}
\begin{figure}[h]
    \hspace{-0.5cm} 
    \includegraphics[scale = 0.8]{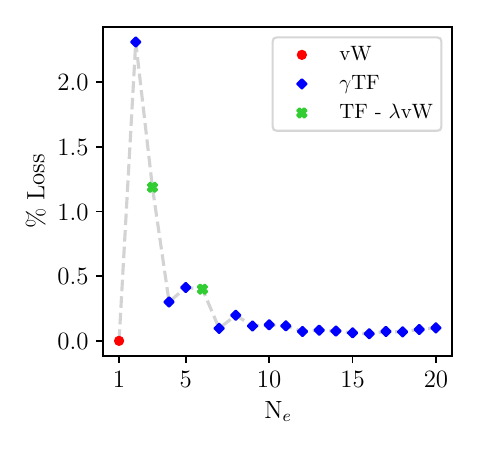}
    \caption{Percentage error of the best-performing functionals found by the SR against the total number of electrons, $N_e$. 
    In this case the vW and TF functional are included as fundamental functions (together with the electron density and
    its first and second derivative) as SR inputs. The vW functional is found for the single-electron case, while the TF one is 
    the best approximation at almost any other electron number. Occasionally, a linear combination of the TF and vW functionals 
    is found to be the best performing kinetic-energy density. Here, $\gamma$ represents a scaling factor.
    }
    \label{fig:vWTFSearch}
\end{figure}

In this search the vW and TF functionals (with their exact prefactors 1/8 and $\pi^{2}/6$, respectively) are included in the 
input set, in addition to those functions utilised in the previous unconstrained search. The idea is to give to the genetic 
search more key ingredients, so that it spends less time in discovering the basic functionals, while also attributing less complexity 
penalty to their presence. The hope is that this would enable more complex functionals to be explored. Unfortunately, this
expectation has not been met. In fact, while some unusual functionals has appeared during the search, these are 
consistently ``outperformed'' (in the context of the performance-complexity trade-off) by either the vW, TF, or the combined TF-vW 
functionals at every electron count. The summary of these findings is presented in Fig.~\ref{fig:vWTFSearch}. 

\begin{figure}[h]
    \hspace{-0.5cm} 
    \includegraphics[scale = 0.8]{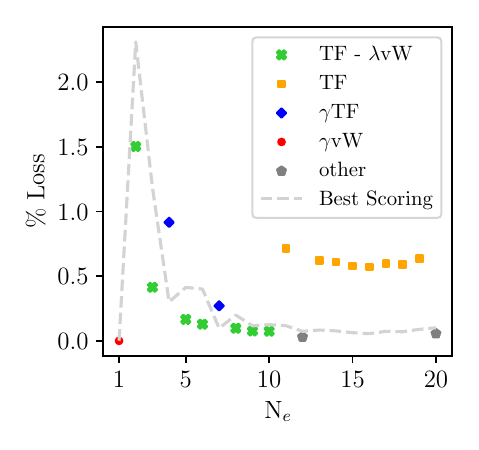}
    \caption{Percentage loss error of the second best performing functionals against the number of electrons in the system, $N_e$. 
    In this search the vW and TF functional are included as fundamental functions (together with the electron density and its 
    first and second derivative) as SR inputs. The grey dotted line indicates the best performing equations found in this search. 
    The second best equations are frequently better performing KEDFs, at the expense of increased functional complexity. 
    Here, $\gamma$ represents a scaling factor.
     }
    \label{fig:vWTF2BSearch}
\end{figure}

As expected, the vW functional tops the search for the one-electron case, this time with essentially zero percentage loss.
Then, at any other electron count the TF expression appears to be at the Pareto front, namely it is the best trade-off between 
accuracy and complexity. However, in a few cases this is outperformed by a linear combination of vW and TF functional,
which are more complex, but seem to have the best percentage improvement against the next lower complexity level. 
This observation suggests that it may be a good idea to look at the second-best expressions found by the SR, namely
at those functionals that rank second in their score [see Fig.~\ref{fig:Pareto}(b)]. Such exercise is performed in 
Fig.~\ref{fig:vWTF2BSearch}, revealing that there are reasonably better performing functionals found at many of the 
different $N_e$. These were previously ignored, since the Pareto front analysis did determine that the performance 
improvement is insufficient to merit the increase in equation complexity. This feature is an inevitable consequence of 
running a multi-objective optimisation. In any case, we find that for intermediate electron counts a linear combination of 
vW and TF functional systematically improves the loss, while for large $N_e$ the second-best function has no longer
a vW or TF form.

\begin{figure}[h]
    \hspace{-0.5cm} 
    \includegraphics[scale = 0.8]{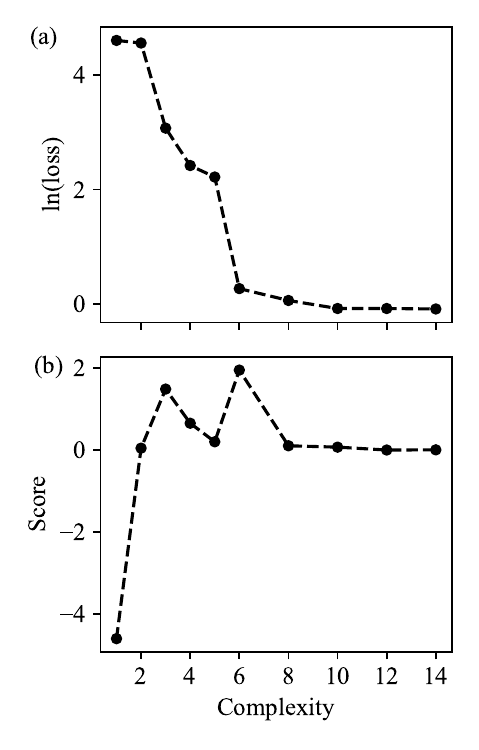}
    \caption{Demonstration of the ambiguity in the identification of the Pareto front during the three electron 
    unconstrained SR search. The score metric, as described in the text, is in panel (b). Peaks coincide 
    with cliffs in the loss curve, indicating the greatest improvement in predictive power for the smallest increase 
    in complexity (in the spirit of Occam's Razor). In this case, two peaks, admitting almost the same height emerge 
    in the score profile for complexity 3 and 6, respectively. This underlines where the differences in Fig.~\ref{fig:vWTFSearch} 
    and Fig.~\ref{fig:vWTF2BSearch} emerge from.}
    \label{fig:PFP}
\end{figure}

The sometimes counter-intuitive ordering of the first and second best-performing KEDF forms at a given complexity is due 
to the search failing to yield a Pareto front with a single, distinct cliff edge in the log(loss) against complexity analysis (see 
Fig.~\ref{fig:PFP}). In some cases, in fact, two cliff edges emerge, indicating separate significant improvements in the loss.
The SymbolicRegression.jl library employs a ``parsimony'' factor that penalizes complexity, by combining the loss with a
measure of frequency and recency of expressions occurring at a given complexity~\cite{Cranmer2023InterpretableML}. 
When two prominent cliff edges appear, this factor can slightly bias the results towards either simpler or more complex 
expressions. All throughout this work we have used chosen a parsimony setting that favors more complex expressions, 
thus de-emphasizing the dramatic cliff edge associated with the genetic search discovering the well-established TF functional.

Going back to the results of Fig.~\ref{fig:vWTF2BSearch}, it is worth noting that the $\lambda$ prefactor in the expressions 
containing a linear combinations of the form `TF - $\lambda$vW' is found to have an average of $0.26545$, with a relatively 
small standard deviation of $0.007658$ between systems of differing numbers of electron (although this is rather statistically 
insignificant). This is found to admit a sign different from that of the three-dimensional case, where the optimal $\lambda$ in 
a $\text{TF}+\lambda\text{vW}$ model is numerically determined to be $\approx 0.2$ \cite{Yonei1965OnTW}. As for the 
$\gamma$TF expressions, the $\gamma$ multiplicative factor is consistently smaller than one, corresponding to the unconstrained 
search results for the TF multiplicative constant being consistently less than the analytical value.

\subsection{Enhancement Factor Search}
The uniform scaling relation expressed in Eq.~(\ref{eq:Scaling}) is a constraint, which can be encoded directly into the 
search. By framing the search such that dimensional consistency is enforced, the SR process does not waste resources 
to compute expressions, which could never correspond to a true KEDF~\cite{Wang2022SymbolicRI,Udrescu2019AIFA}. 
In order to accommodate such constraint, we rewrite the functional as
\begin{equation}
    T_\text{s}[\rho] = \frac{\pi^2}{6}\int [\rho(x)]^3 F(s(x), q(x), k(x))\:, \label{eq:EF}\:
\end{equation}
where, $s(x)$ is given by
\begin{equation}
    s(x) = \frac{d \rho(x)}{dx}\frac{1}{[\rho(x)]^2}\:, \label{eq:s}
\end{equation}
$q(x)$ is
\begin{equation}
    q(x) = \rho(x)\frac{d^2 \rho(x)}{dx^2}\frac{1}{[\rho(x)]^4}\:, \label{eq:q}
\end{equation}
and $k(x)$ writes
\begin{equation}
    k(x) = \rho(x)\frac{d^2 \rho(x)}{dx^2}\frac{1}{[\frac{d \rho(x)}{dx}]^2}\:. \label{eq:k}
\end{equation}
The quantities $s(x)$, $q(x)$, and $k(x)$ are all dimensionless, and are now provided to the SR 
as input functions. $F(s(x), q(x), k(x))$ is known as an enhancement factor, and by forcing the 
SR to learn only this part, the search is significantly narrowed~\cite{Ma2022}. The 
three-dimensional equivalents of $s(x)$ and $q(x)$ are known as the reduced density gradient, 
and the reduce density laplacian respectively. Both are used to detect density inhomogeneity, 
and can be used to detect when a bonding region is being considered~\cite{Pokharel2022ExactCA}. 

\begin{figure}[h]
    \hspace{-0.5cm} 
    \includegraphics[scale = 0.8]{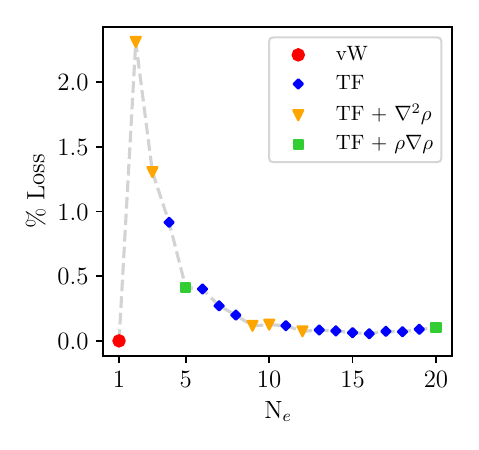}
    \caption{Percentage loss error of the best performing functionals against the number of electrons in 
    the system, $N_e$. In this search the SR is forced to learn the enhancement factor of Eq.~(\ref{eq:EF}), 
    and can make use of the dimensionless functions $s(x)$, $q(x)$ and $k(x)$, respectively in Eq.~(\ref{eq:s}), 
    Eq.~(\ref{eq:q}) and Eq. (\ref{eq:k}). The grey dashed line os to guide the eye.
    }
    \label{fig:EF}
\end{figure}
Our results for this last search are presented in Fig.~\ref{fig:EF}. The functionals at the Pareto front, given in the legend, 
are the result of multiplying the enhancement factor with the TF prefactor. The vW kinetic-energy density is found again for 
the one-electron case through an expression involving $[s(x)]^2$. The multiplicative constant is again close to the theoretical 
value, and the expression yields an overall percentage loss of 0.00027\%. The rest of the expressions are variants of the TF
functional, with no especially interesting additional terms. The integral of the $\rho (d\rho(x)/dx)$ and $d^2 \rho(x)/dx^2$ 
terms range from \(10^{-14}\) to \(10^{-11}\); essentially zero, although the presence of these terms in self-consistent calculations 
have been shown to reproduce shell structure in a three dimensional setting \cite{Yang1986VariousFF}.


\section{Conclusions}

In the search for a machine-learning expression for the non-interacting kinetic-energy density, which admits stable
functional derivatives, we have performed symbolic regression for the one-dimensional case as a function of the 
number of electrons. In particular, we have pursued different strategies, ranging from a completely uninformed search 
to an attempt at learning the Thomas-Fermi enhancement factor. In all searches, for one electron, the symbolic regression 
has been able to converge to the exact solution, namely the vW functional. In contrast, for many-electrons a semi-local 
SR appeared not to be sufficient, in particular when quantum mechanical effects are most pronounced, in the few-electron 
case. By analysing the best- and second-best performing functionals as determined by the SR, we have demonstrated that
sometimes a counter-intuitive balance exists between performance and complexity. 

Despite the fact that this work is only meant to be a first experiment in the use of symbolic regression for the search of a KEDF,
it opens several new avenues. One option may be to consider a determined non-local contribution and relearn the semi-local 
component. This strategy was followed in Ref.~\cite{Ma2022} in the SR search for the exchange-correlation functional. Ideally, 
one may then try to learn both the non-local and semi-local components. While this task requires a significant increase in the 
computational demand, the adaptations of the SymbolicRegression.jl library performed in this work to make it 
compatible with functionals lay the groundwork for further expansions. 

When moving to 3D, more performant integration schemes must be employed. Furthermore, in order to incorporate known 
functional constraints, such as the asymptotic constraints at infinity, a straightforward post-processing step mapping the SR 
expression to the desired output domain will force an inductive bias on the model 
\cite{Pokharel2022ExactCA,Nagai2021MachinelearningbasedEC}. Finally, turning our attention to the SR foundations, recent 
steps towards leveraging the 1-Wasserstein distance to quantify the similarity between mathematical 
expressions~\cite{Meznar2024} may introduce a structure that can be exploited for improved convergence.

In conclusion, this work has articulated the nuance of working with SR as a method, and has provided a brute-force investigation 
into the capacity of semi-local analytic functionals in accurately the one dimensional KEDF.
 
\begin{acknowledgments}
This work has been sponsored by Taighde \'Eireann - Research Ireland through the
PhD scholarship program (MAJM), the Advanced Laureate Award (IRCLA/2019/127)
(TDAF) and the AMBER Research Center (12/RC/2278$_-$P2).
We acknowledge Trinity IT Research for the provision of computational 
resources and the Nvidia Academic Hardware grant for the use of GPUs.
\end{acknowledgments}

\end{document}